\documentstyle[prl,floats,aps,twocolumn,epsf,graphicx]{revtex}
\begin{document}
\twocolumn[\hsize\textwidth\columnwidth\hsize\csname
@twocolumnfalse\endcsname

\title{Fermionic Quasinormal Spectrum of the Kerr Black Hole}
\author{Shahar Hod}
\address{The Ruppin Academic Center, Emeq Hefer 40250, Israel}
\date{\today}
\maketitle

\begin{abstract}

\ \ \ We study {\it analytically} the asymptotic quasinormal spectrum of fermionic fields in the Kerr spacetime. 
We find an analytic expression for these black-hole resonances in terms of the black-hole physical 
parameters: its Bekenstein-Hawking temperature $T_{BH}$, 
and its horizon's angular velocity $\Omega$, which is valid in 
the asymptotic limit $1 \ll \omega_I \ll \omega_R$. 
It is shown that according to Bohr's correspondence principle, 
the emission of a  Rarita-Schwinger quantum ($s=3/2$) corresponds to  a fundamental 
black-hole area change $\Delta A=4\hbar \ln 2$, while the emission of a Weyl neutrino field ($s=1/2$) 
corresponds to an adiabatic quantum transition with $\Delta A=0$.
\end{abstract}
\bigskip

]

Gravitational waves emitted by a perturbed black hole are dominated by
`quasinormal ringing', damped oscillations with a {\it discrete}
spectrum (see e.g., \cite{Nollert1} for a detailed review). 
At late times, all perturbations are
radiated away in a manner reminiscent of the last pure dying tones of
a ringing bell \cite{Press,Cruz,Vish,Davis}. 
Being the characteristic `sound' of
the black hole itself, these free oscillations 
are of great importance from 
the astrophysical point of view. They allow a direct way of identifying the spacetime 
parameters (especially, the mass and angular momentum of the black hole). 
This has motivated a flurry of activity with the aim of computing the spectrum of black-hole 
oscillations.

It turns out that for a Schwarzschild black hole, for a given angular harmonic index $l$ 
there exist an infinite number of quasinormal modes, for $n=0,1,2,\dots$, characterizing oscillations with decreasing
relaxation times (increasing imaginary part) \cite{Leaver,Bach}. On 
the other hand, the real part of the Schwarzschild black-hole frequencies approaches an asymptotic 
{\it constant} value \cite{Nollert2,Andersson}.

The quasinormal modes (QNMs) have been the subject of much recent attention 
(see \cite{HodKesh} for a detailed list of 
references), with the hope that these classical resonances 
may shed some light on the elusive theory of quantum gravity. 
These recent studies are motivated by an earlier work \cite{Hod1}, in which 
a possible correspondence between the black-hole classical resonances and the quantum properties of its 
surface area was suggested.

The quantization of black holes was proposed long ago by Bekenstein
\cite{Beken1,Beken2}, based on the remarkable observation that the horizon
area of a non-extremal black hole behaves as a classical 
adiabatic invariant. In the spirit of the Ehrenfest principle
\cite{Ehren} -- any classical adiabatic invariant
corresponds to a quantum entity with a discrete spectrum, and based on the idea of a minimal increase in 
black-hole surface area \cite{Beken1}, it was conjectured that the horizon area of a quantum
black hole should have a discrete spectrum of the form

\begin{equation}\label{Eq1}
A_n=\gamma {\ell^2_P} \cdot n\ \ \ ;\ \ \ n=1,2,3,\ldots\ \  ,
\end{equation}
where $\gamma$ is a dimensionless constant, and 
$\ell_P=(G\hbar/c^3)^{1/2}$ is the
Planck length (we use units in which $G=c=\hbar=1$ henceforth). 
This type of area quantization has since been reproduced based on various other considerations 
(see e.g., \cite{HodA} for a detailed list of references). 

In order to determine the value of the coefficient $\gamma$, 
Mukhanov and Bekenstein \cite{Muk,BekMuk,Beken3} have suggested, 
in the spirit of the Boltzmann-Einstein formula in statistical physics, to 
relate $g_n \equiv \exp[S_{BH}(n)]$ to the number of the black hole microstates 
that correspond to a particular external macro-state, where $S_{BH}$ is the black-hole entropy. 
In other words, $g_n$ is the degeneracy of the $n$th area eigenvalue. 
Now, the thermodynamic relation between black-hole surface area and entropy, $S_{BH}=A/4\hbar$, 
can be met with the requirement that $g_n$ has to be an integer for every $n$ only when
 
\begin{equation}\label{Eq2}
\gamma =4\ln{k} \  ,
\end{equation}
where $k$ is some natural number. 

Identifying the value of $k$ requires further input. This information 
may emerge by applying {\it Bohr's correspondence principle} to 
the (discrete) quasinormal mode spectrum of black holes. 
Based on the correspondence principle, it was argued \cite{Hod1} that 
the asymptotic resonances are given by 
(we assume a time dependence of the form $e^{-i\omega t}$),

\begin{equation}\label{Eq3}
\omega = T^{s}_{BH}\ln3 -i 2\pi T^{s}_{BH} (n+{1\over2})\  ,
\end{equation}
where $T^{s}_{BH}=1/8\pi M$ is the Bekenstein-Hawking temperature of the Schwarzschild black hole. 
An analytical proof of this equality was later given in \cite{Motl}. 

The emission of a quantum of frequency $\omega$ results in a change $\Delta M=\hbar \omega_R$ in 
the black-hole mass. Assuming that $\omega$ corresponds to the asymptotic 
limit Eq. (\ref{Eq3}), and using the first-law of black-hole thermodynamics 
$\Delta M={1 \over 4}T^{s}_{BH} \Delta A$, this implies 
a change $\Delta A=4\hbar \ln3$ in the black hole surface area. 
Remarkably, this value is in accord with the Bekenstein-Mukhanov prediction, 
Eq. (\ref{Eq2}).

The possible correspondence between the black-hole classical resonances and the quantum 
properties of its surface area has triggered a flurry of research 
attempting to calculate the asymptotic ringing frequencies of various types of 
black holes (for a detailed list of references see, e.g., \cite{HodKesh}). 
For instance, the asymptotic QNM spectrum of the 
Reissner-Nordstr\"om (RN) black hole was calculated in \cite{MotNei}. This was followed by 
an analytical calculation of the asymptotic QNM frequencies of a {\it charged} field in the 
RN spacetime \cite{HodR}. The results presented in \cite{MotNei,HodR} 
indicate that the asymptotic resonances of a charged field correspond to a 
fundamental black-hole area change of $\Delta A=4\hbar \ln 2$, 
in accord with the general prediction Eq. (\ref{Eq2}).

It should be emphasized, however, that less is known about the corresponding 
QNM spectrum of the generic (rotating) Kerr black hole, which is the most interesting 
from a physical point of view. 
Former studies of the Kerr asymptotic spectrum \cite{Leaver,Det,Ono,Ber1,Ber2,Ber3} used numerical tools, 
and there are in fact no analytical results for the asymptotic QNM frequencies of the Kerr black hole. 
In this work we provide analytical formulae for the Kerr QNM spectrum in the 
asymptotic limit $1 \ll \omega_I \ll \omega_R$. 
We consider fermionic modes, characterized by half-integer spins.

The dynamics of fermionic fields in the Kerr spacetime is governed 
by the Teukolsky equation \cite{Teukolsky}

\begin{equation}\label{Eq4}
\Delta {{d^2R_{lm}} \over {dr^2}} +(s+1)(2r-2M){{dR_{lm}} \over {dr}} +V(r)R_{lm}=0\  ,
\end{equation}
with

\begin{eqnarray}\label{Eq5}
V(r;a,\omega,s,l,m)&=&\Delta^{-1}[(r^2+a^2)^2{\omega}^2-2ma\omega r+a^2m^2]\nonumber \\
&&+is\Delta^{-1}[ma(2r-2M)-\omega(r^2-a^2)]\nonumber \\
&&+2is\omega r-{(a\omega)}^2-A_{slm}\  ,
\end{eqnarray}
where $\Delta \equiv (r-r_+)(r-r_-)$ [$r_{\pm} =M \pm (M^2-a^2)^{1/2}$ are the 
black hole (event and inner) horizons], and 
$a\equiv J/M$ is the black hole angular momentum per unit mass. 
The field spin-weight parameter $s$ takes the values $\pm 1/2$ and $\pm 3/2$ for a 
two-component Weyl neutrino field, and for a Rarita-Schwinger field, respectively. 
The angular separation constants $A_{slm}(a\omega)$ are determined from an independent 
differential equation \cite{Leaver}, which governs the angular dependence of the field.

The black hole QNMs correspond to solutions of the wave equation with the physical boundary
conditions of purely outgoing waves at spatial infinity 
and purely ingoing waves crossing the event horizon \cite{Detwe}. Such boundary 
conditions single out a {\it discrete} set of resonances $\{\omega_n\}$. 
The solution to the radial Teukolsky equation may be expressed as \cite{Leaver} 
(assuming an azimuthal dependence of the from $e^{im\phi}$)

\begin{eqnarray}\label{Eq6}
R_{lm}& = &e^{i\omega r} (r-r_-)^{-1-s+i2M\omega+i\sigma_+} (r-r_+)^{-s-i\sigma_+}\nonumber \\
&&\Sigma_{n=0}^{\infty} d_n \Big({{r-r_+} \over {r-r_-}}\Big)^n\  ,
\end{eqnarray}
where $\sigma_{+} \equiv (\omega r_{+}-ma)/(r_{+}-r_{-})$. 

The sequence of expansion coefficients $\{d_n:n=1,2,\ldots\}$ is determined by a 
recurrence relation of the form \cite{Leaver}

\begin{equation}\label{Eq7}
\alpha_n d_{n+1}+\beta_n d_n +\gamma_n d_{n-1}=0\  ,
\end{equation}
with initial conditions $d_0=1$ and $\alpha_0 d_1+\beta_0 d_0=0$.
The recursion coefficients $\alpha_n,\beta_n$, and $\gamma_n$ are given in \cite{Leaver}. The quasinormal frequencies are determined by the requirement that the series in Eq. (\ref{Eq6}) 
is convergent, that is $\Sigma d_n$ exists and is finite \cite{Leaver}.

The physical content of the recursion coefficients becomes clear when they are expressed 
in terms of the black-hole physical parameters \cite{HodKesh}: the Bekenstein-Hawking temperature $T_{BH}=(r_{+}-r_{-})/A$, 
and the horizon's angular velocity $\Omega=4\pi a/A$, where $A=4\pi(r_+^2+a^2)$ is the black-hole surface area. 
The recursion coefficients obtain a simple form in terms of these physical quantities, 

\begin{equation}\label{Eq8}
\alpha_n=(n+1)(n+1-s-2i\beta_{+}\hat\omega)\  ,
\end{equation}

\begin{eqnarray}\label{Eq9}
\beta_n& = &-2(n+{1 \over 2}-2i\beta_{+}\hat\omega)
(n+{1 \over 2}-2i\omega r_{+})\nonumber \\
&&-\lambda_{slm}-s-{1 \over 2}\  ,
\end{eqnarray}
and
\begin{equation}\label{Eq10}
\gamma_n=(n-4iM\omega)(n+s-2i\beta_{+}\hat\omega)\  ,
\end{equation} 
where $\beta_{+} \equiv (4\pi T_{BH})^{-1}$ is the black-hole inverse temperature, 
$\hat\omega \equiv \omega-m\Omega$, 
and $\lambda_{slm}\equiv A_{slm}+{(a\omega)}^2-2ma\omega$. 

The angular separation constants $A_{slm}(a\omega)$ are determined 
by an independent recurrence relation \cite{Leaver}. 
The asymptotic large $a\omega_R$ limit of the angular separation constants 
is given by \cite{Abr,Flam,Casa,BCC}

\begin{equation}\label{Eq11}
A_{slm}=-(a\omega)^2+2q^{(1)}_{slm}a\omega+q^{(0)}_{slm}+q^{(-1)}_{slm}(a\omega)^{-1}+\cdots \  ,
\end{equation} 
where the expansion coefficients $\{q^{(n)}_{slm}\}$ are given in \cite{Casa,BCC}. 
Remarkably, one finds 

\begin{equation}\label{Eq12}
A_{slm}=-(a\omega)^2+2ma\omega+O(1)\  ,
\end{equation} 
for fermionic modes characterized by $l=m+|s|-{1 \over 2}$, 
thus yielding $\lambda_{slm}=O(1)$ in these cases.

The Teukolsky equation also describes the propagation of fermionic fields 
in the RN spacetime \cite{HodKeshrnn} [one should 
simply replace $r_{\pm}=M \pm (M^2-a^2)^{1/2}$ by 
$r_{\pm}=M \pm (M^2-Q^2)^{1/2}$, and take $a=0$ elsewhere. In addition, 
$A_{slm}=l(l+1)-s(s+1)$ for the spherically symmetric RN spacetime \cite{Leaver}.] 
The spectrum of the RN QNMs may be found using a recursion relation of the form Eq. (\ref{Eq7}). 
The $\{\alpha_n\}$ and $\{\gamma_n\}$ coefficients of 
the RN perturbations have exactly the same form as in the Kerr case 
(where for the RN black hole $\hat\omega=\omega$). 
In addition, one finds $\beta^{RN}_n=\beta^{Kerr}_n+\lambda_{slm}(a\omega)$. 
Taking cognizance of the asymptotic limit Eq. (\ref{Eq12}), one finds that for fermionic modes 
with $l=m+|s|-{1 \over 2}$ the $\{\beta_n\}$ coefficients of the Kerr black hole coincide 
with those of the RN black hole (in addition to the coincidence of the $\{\alpha_n\}$ and $\{\gamma_n\}$ terms).

Thus, one finds that the asymptotic quasinormal frequencies 
of fermionic fields (with $l=m+|s|-{1 \over 2})$ 
in the Kerr spacetime are related to the corresponding asymptotic frequencies of the RN black hole. 
The asymptotic $1 \ll \omega_I$ behavior of the RN resonances is determined by the equation \cite{MotNei,Note2}

\begin{equation}\label{Eq13}
e^{-4\pi\beta_{+}\omega}=1\ ,
\end{equation}
for the two-component Weyl field, and by \cite{MotNei,Note3}

\begin{equation}\label{Eq14}
2e^{-4\pi\beta_{+}\omega}+3e^{-8\pi M\omega}=-1\  ,
\end{equation}
for the Rarita-Schwinger field.

The preceding discussion indicates that expressions similar to Eqs. (\ref{Eq13}) and 
(\ref{Eq14}) must hold true for the asymptotic QNMs 
of fermionic fields (with $l=m+|s|-{1 \over 2}$) in the Kerr black-hole spacetime. 
Equations (\ref{Eq13}) and (\ref{Eq14}) suggest that the spectrum depends 
on the combinations $\beta_+ \omega$ and $M\omega$ appearing in Eqs. (\ref{Eq8})-(\ref{Eq10}), 
but does not depend explicitly on $\omega r_+$. 
From Eqs. (\ref{Eq8})-(\ref{Eq10}) one learns that the analogy 
between the asymptotic spectrum of fermionic fields 
in the Kerr spacetime, and the corresponding spectrum in the RN spacetime 
is obtained by applying the transformation 
$\beta_{+}\omega \to \beta_{+}\hat\omega$ in Eqs. (\ref{Eq13}) and (\ref{Eq14}). 
Using this transformation, one finds that the asymptotic $1 \ll \omega_I \ll \omega_R$ 
quasinormal spectrum of neutrino modes with $l=m$ is determined by the equation

\begin{equation}\label{Eq15}
e^{-4\pi\beta_{+}(\omega-m\Omega)}=1\  ,
\end{equation}
thus yielding

\begin{equation}\label{Eq16}
\omega = m\Omega-i 2\pi T_{BH} n\  .
\end{equation}
The corresponding spectrum of a Rarita-Schwinger field with $l=m+1$ is given by 

\begin{equation}\label{Eq17}
2e^{-4\pi\beta_{+}(\omega-m\Omega)}+3e^{-8\pi M\omega}=-1\  ,
\end{equation}
thus yielding 

\begin{equation}\label{Eq18}
\omega=m\Omega +T_{BH}\ln 2-i 2\pi T_{BH} (n+{1 \over 2})\  .
\end{equation}

The emission of a quantum of frequency $\omega$ and azimuthal number $m$ 
results in a change $\Delta M=\hbar \omega_R$ in the black-hole mass, 
and a change $\Delta J=m\hbar$ in its angular momentum. 
Substituting the fundamental resonances, Eqs. (\ref{Eq17}) and (\ref{Eq18}), into 
the first law of black-hole thermodynamics 

\begin{equation}\label{Eq19}
\Delta M={1 \over 4}T_{BH} \Delta A + \Omega \Delta J\  ,
\end{equation}
one obtains 

\begin{equation}\label{Eq20}
\Delta A =0\  ,
\end{equation}
for the Weyl neutrino field, and 

\begin{equation}\label{Eq21}
\Delta A =4\hbar \ln 2\  ,
\end{equation}
for the Rarita-Schwinger field. 

Thus, the emission of a neutrino quantum (with $l=m$ and $m\gg 1$) 
from a Kerr black hole corresponds to an {\it adiabatic} transition, for which 
there is no net change in the black-hole surface area 
(though the mass and angular-momentum of the black hole have been 
changed by the emission). We note that, remarkably this is a generic feature of 
asymptotic neutrino resonances: they 
correspond to $\Delta A=0$ for both the Schwarzschild \cite{Motl}, RN \cite{HodKeshrnn}, 
and Kerr black holes. 

The emission of a Rarita-Schwinger fermion (with $l=m+1$ and $m\gg 1$) 
results with a fundamental change $\Delta A=4\hbar \ln 2$ in black-hole surface 
area. Remarkably, this fundamental change in the surface area 
is in accord with the Bekenstein-Mukhanov general prediction, Eq. (\ref{Eq2}). 

In summary, motivated by novel results in the theory of black-hole quantization, 
we have studied analytically the QNM spectrum of fermionic fields 
in the Kerr black-hole spacetime. 
It was shown that in the asymptotic limit $1 \ll \omega_I \ll \omega_R$, these black-hole resonances 
can be expressed in terms of the black-hole physical parameters: 
its temperature $T_{BH}$, and its angular velocity $\Omega$. 

The analysis of the Kerr quasinormal spectrum enabled us to test the applicability of 
Bohr's correspondence principle to the 
quantization of black holes in generalized situations, in 
which there is no one-to-one correspondence between the energy of the emitted quantum and the resulting 
change in black-hole surface area [The Kerr black hole has a chemical potential, in the 
form of the black-hole angular velocity $\Omega$, see Eq. (\ref{Eq19}).] 
We have shown that according to the correspondence principle, the emission of a Rarita-Schwinger 
quantum characterized by $l=m+1$ 
induces a fundamental change in the Kerr black-hole surface area, $\Delta A=4\hbar\ln2$. 
Remarkably, this area unit is universal in the sense that it is {\it independent} of the black-hole 
parameters. The emission of a Weyl neutrino with $l=m$ corresponds to an adiabatic transition, in which there 
is no change in the black-hole surface area.

\bigskip
\noindent
{\bf ACKNOWLEDGMENTS}
\bigskip

The research of SH was supported by G.I.F. Foundation. 
I thank Uri Keshet for numerous discussions, as well as for 
a continuing stimulating collaboration.

\end{document}